# Experimental Demonstration of Imperfection-Agnostic Local Learning Rules on Photonic Neural Networks with Mach-Zehnder Interferometric Meshes

**Luis El Srouji, Mehmet Berkay On, Yun-Jhu Lee, Mahmoud Abdelghany, S. J. Ben Yoo**
*University of California, Davis*
Author e-mail address: lzelsrouji@ucdavis.edu, sbyoo@ucdavis.edu

**Abstract:** Mach-Zehnder Interferometric meshes are attractive for low-loss photonic matrix multiplication but are challenging to program. Using least-squares optimization of directional derivatives, we experimentally demonstrate that desired matrix updates can be implemented agnostic to hardware imperfections. © 2024 The Author(s)

## 1. Introduction

Integrated photonics are attractive for analog computation due to higher transmission speeds and larger bandwidths than traditional electronics [1]. Photonic accelerators have targeted matrix multiplication in particular because the quadratic computational complexity of the operation can slow computationally intensive applications like neural network inference [2]. Various photonic devices can be arranged into analog photonic matrix multipliers to take advantage of the higher throughput and operating speeds, including many architectures of Mach-Zehnder Interferometric (MZI) meshes [3–5].

   Despite these advantages, programming MZI meshes is challenging because of the nonlinear relationship between the tunable parameters available (phase shifts) and the resultant matrix implemented. Multiple algorithms have been proposed to implement a target matrix [3, 4], along with gradient-descent-based methods for training neural networks [6, 7]. In the latter set of algorithms, the target matrix is unknown in advance, and the parameters are optimized according to gradient descent of some cost or loss function. These methods are only applicable when a loss function is well-defined and differentiable.

   In neuromorphic computing, however, brain-inspired neural networks contain many recurrent structures whose dynamics are not well-defined by a loss function and may not be easily differentiable. As such, brain-like learning rules have been proposed that can be used to train these neural networks with only the information local to a given synaptic connection [8–10]—in other words, using only the activity of the sending and receiving neurons. These local learning rules dictate the change to a single synaptic weight, but this is problematic for MZI meshes because of the aforementioned nonlinear mapping.

   We propose and experimentally demonstrate a scheme for implementing local learning rules on MZI meshes. The implementation is agnostic to fabrication variations, thermal crosstalk, and requires no calibration.

## 2. Local Learning by Central-Difference Approximation

Each MZI implements an arbitrary 2x2 unitary operation that can be programmed to control the power coupling and phase difference of signals between each of its input and output ports. In the phasor domain, the coupling of electromagnetic waves between input and output ports can be written as the transfer matrix below. $\vec{A}$ and $\vec{B}$ denote the phasor forms of the optical fields at the input and output ports, respectively, and $\phi$ and $\theta$ represent phase shifts within a single functional MZI unit (as shown in Fig. 1a):

$$\vec{B} = \begin{bmatrix} e^{j\phi}\sin(\theta/2) & \cos(\theta/2) \\ e^{j\phi}\cos(\theta/2) & -\sin(\theta/2) \end{bmatrix} \vec{A} \qquad (1)$$

Meshes of MZIs can be used to implement larger matrix operations, and because the operation implemented is unitary, this multiplier is nearly lossless, allowing for low-power accelerator designs. The MZI mesh can be operated coherently or incoherently depending on whether the inputs are generated by a single laser or multiple lasers. Because the MZI mesh is attractive in low-power applications, it is well-suited for use in neuromorphic accelerators. Within neuromorphic computing, spiking neurons are preferred for low-power signal encodings and as such an incoherent mode of operation is sufficient. In addition, the synaptic weighting in the human brain can be represented with positive magnitudes only, with the distinction between excitatory versus inhibitory synapses being a result of the neurotransmitter used; this means the synapse never changes sign and can be implemented by the interpretation of each output port on the incoherently excited MZI mesh.

   A given weight matrix implemented by an MZI mesh can be represented as a multivariable function $W(\Theta): \mathbb{R}^n \rightarrow \mathbb{R}^m$ where $\Theta$ represents the set of tunable parameters, and $W$ is the set of weights implemented. The mapping is nonlinear and sensitive to hardware-related impairments, making it intractable to calculate the derivative analytically. Instead, a multi-variate central difference approximation can be used to calculate the directional derivative of each weight, $\nabla_{\vec{v}} W(\Theta)$, with respect to a particular parameter update vector, $\vec{v}$:

$$\nabla_{\vec{v}} W(\Theta) = \frac{W(\Theta + h\vec{v}) - W(\Theta + h\vec{v})}{2h||\vec{v}||} \quad (2)$$

A single directional derivative may not be well-aligned with the target update given by a local learning rule, $\Delta W$. Thus, multiple randomly selected directional derivatives are needed to poll the space. To ensure that the implemented step is in the direction of $\Delta W$, we implement a least-squares optimization to find the linear combination of directional derivatives that is most closely aligned. Next, the true weight update, $\Delta \widehat{W}$, is calculated and subtracted from the target. The algorithm can be iterated to some desired metric of convergence, such as a threshold $\ell^2$-norm of the target difference vector, $\Delta W$. The pseudo code for this algorithm is shown below:

---
*Algorithm 1: Local Learning by Central-Difference Approximation*
---

1) Calculate desired weight matrix update, $\Delta W$, according to local learning rule.
2) Calculate several directional derivatives, $\nabla_{\vec{v}} W(\Theta)$, for the weight matrix with respect to all tunable elements:
   $\vec{x}_k = \nabla_{\vec{v}_k} W(\Theta)$ (as a single matrix $X = [\vec{x}_1 \ ... \ \vec{x}_k]$)
3) Find the linear combination of $\vec{x}_k$ that minimizes the difference with the desired update:
   $\min_{\vec{a}} \{(\Delta W - X\vec{a})^2\} \rightarrow \vec{a} = (X^T X)^{-1} X \Delta W$
4) Apply the update according to the closest linear combination of directional derivatives:
   $\Theta \rightarrow \Theta + V\vec{a}$ where $V = [\vec{v}_1 \ ... \ \vec{v}_k]$
5) Measure the true weight update, $\Delta \widehat{W}$, and subtract from the target. Return to step 2) and iterate to convergence.
   $\Delta W \rightarrow \Delta W - \Delta \widehat{W}$

## 3. Experimental Demonstration

The above algorithm was experimentally demonstrated using a 4x4 subset of a 6×6 MZI mesh fabricated on the silicon photonic multi-project wafer run by AIM Photonic foundry. The device consists of a rectangular MZI mesh configuration [4] with thermo-optic phase shifters as the tuning elements. On-chip SiGe photodetectors monitor the optical power at the input and output ports of the MZI mesh. An external computer implemented the algorithm and communicated with the MZI device using a digital-to-analog converter (DAC) for phase-shifter control and an analog-to-digital converter (ADC) for reading out the internal photodetectors. For this experiment, only the internal phase shifter $\theta$ of each MZI unit was used for tuning. Fig. 1a) shows a diagram of the experimental setup, and Fig. 1b) shows a photograph of the photonic integrated circuit.

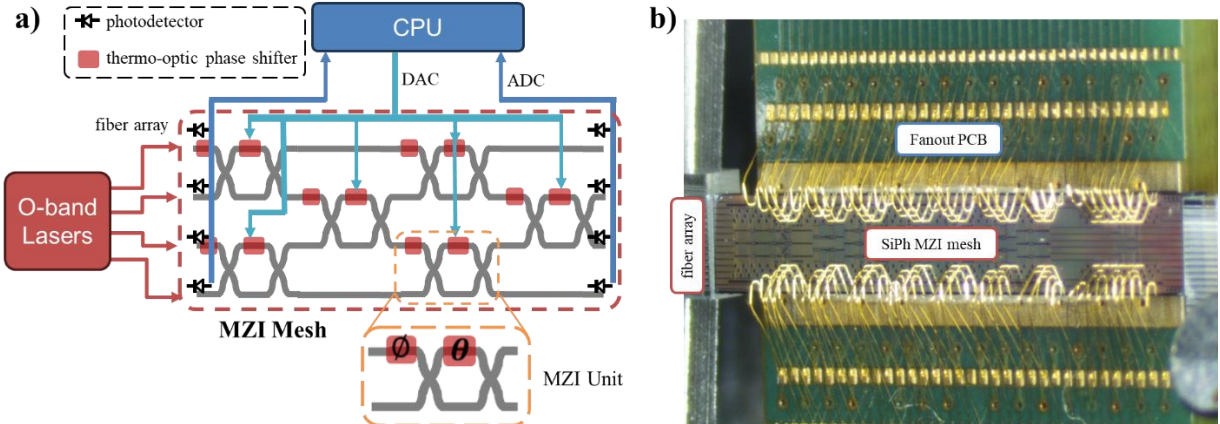

*Fig. 1a) Diagram of experiment. b) Photograph of device.*

To demonstrate the convergence of the proposed algorithm, we randomly initialize the MZI mesh and used Algorithm 1 to implement six different permutation matrices on the mesh. We randomly initialize the MZI mesh for each permutation matrix with each phase shifter biased towards the middle of the output range for our DAC. If a phase shifter is pushed out of the control range of the DAC, it is again randomly re-initialized towards the middle of the controllable range. Fig. 2a) shows the $\ell^2$-norm of the difference vector, $\Delta W$, between the target and current matrix as we iterate Algorithm 1. After 50 iterations, the $\ell^2$-norm of each difference vector falls below 0.07, with some trials reaching less than 0.001 with as few as 14 iterations. This norm corresponds to less than a 1% error on the value of each matrix element. For local learning, however, we do not know the target matrix but instead a target difference vector, $\Delta W$. Fig. 2b) shows a plot of the number of iterations required to converge the $\ell^2$-norm to within 0.08 (representing the same 1% error) compared to the initial magnitude ($\ell^2$-norm) of the difference vector. Because it is common to normalize the size of updates during neural network training, a fixed upper bound on the number of algorithmic iterations can be selected based on the noise floor and precision of the neuromorphic accelerator. Additionally, because local learning rules are applied to all neurons in parallel, an accelerator applying this algorithm can potentially train the neural network in constant algorithmic time regardless of network size.

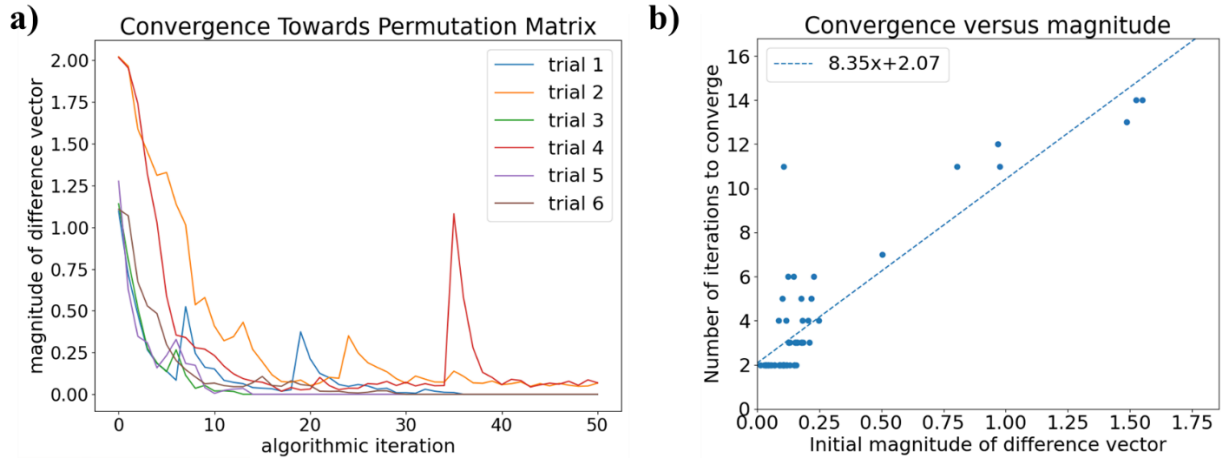

*Fig. 2a) Convergence towards one of six permutation matrices. b) Number of iterations required to implement a change in magnitude*

## 4. Conclusion

In this presentation, we experimentally demonstrate an algorithm for implementing neuromorphic local learning rules on photonic matrix multipliers based on MZI meshes. The algorithm can implement a desired target weight update, $\Delta W$, in a manner agnostic to the underlying mapping between tunable parameters and the matrix implemented.

**This material is based upon work supported by the Air Force Office of Scientific Research (AFOSR) under award number FA9550-22-1-0532 and under award number FA9550-181-1-0186, and by the Office of the Director of National Intelligence, Intelligence Advanced Research Projects Activity under Grant 2021-21090200004.**